\documentclass{Interspeech2024}
\usepackage{booktabs}
\usepackage{arydshln}
\usepackage{balance}

\interspeechcameraready

\title{Exploiting Foundation Models and Speech Enhancement for Parkinson's Disease Detection from Speech in Real-World Operative Conditions}
\name[affiliation={1}]{Moreno}{La Quatra}
\name[affiliation={1}]{Maria Francesca}{Turco}
\name[affiliation={2}]{Torbjørn}{Svendsen}
\name[affiliation={2}]{Giampiero}{Salvi}
\name[affiliation={3}]{Juan Rafael}{Orozco-Arroyave}
\name[affiliation={2,4}]{Sabato Marco}{Siniscalchi}

\address{
  $^1$Kore University of Enna, Italy 
  $^2$Norwegian University of Science and Technology, Norway \\
  $^3$University of Antioquia, Colombia 
  $^4$Università degli Studi di Palermo, Italy}
\email{\{sabatomarco.siniscalchi\}@unipa.it}


\begin{document}

\maketitle

\begin{abstract}

This work is concerned with devising a robust Parkinson's (PD) disease detector from speech in real-world operating conditions using (i) foundational models, and (ii) speech enhancement (SE) methods. To this end, we first fine-tune several foundational-based models on the standard PC-GITA (s-PC-GITA) clean data. Our results demonstrate superior performance to previously proposed models. Second, we assess the generalization capability of the PD models on the extended PC-GITA (e-PC-GITA) recordings, collected in real-world operative conditions, and observe a severe drop in performance moving from ideal to real-world conditions. Third, we align training and testing conditions applaying off-the-shelf SE techniques on e-PC-GITA, and a significant boost in performance is observed only  for the foundational-based models. Finally, combining the two best foundational-based models trained on  s-PC-GITA, namely WavLM Base and Hubert Base, yielded top performance on the enhanced e-PC-GITA.
\end{abstract}

\section{Introduction}
Parkinson’s disease (PD) is a neurodegenerative condition marked by the progressive decline of dopaminergic neurons in the mid-brain~\cite{hornykiewicz1998biochemical, Poewe2017, Rijk2000}. This leads to a range of motor and non-motor symptoms, for instance, tremors, bradykinesia, cognitive deterioration, and  depression~\cite{hornykiewicz1998biochemical,mu2017parkinson, Jankovic2008}. 
During the prodromal stages of PD, patients may exhibit speech disorders, 
with these symptoms emerging as early as five years before the onset of 
significant motor impairments~\cite{pinto2004treatments, Rusz2013}. 
Research indicates a rapid escalation in neuronal damage during the four years following diagnosis~\cite{Rodriguez-Oroz2009}. 
The evaluation of PD through speech analysis has thus gained 
attention among researchers due to its automated, cost-effective, and 
non-intrusive nature, making it a promising method for early PD detection.

In recent years, several works have thus been proposed to detect speech impairments 
and predict PD progression by leveraging solutions based on both shallow and deep models. %
For example, some studies considered models based on Convolutional Neural Networks (CNN) using
spectrograms as input to classify PD patients and Healthy Controls (HCs) or to 
detect dysarthria and predict its severity level~\cite{sonawane2021speech,rios2022end}.
A model combining unidimensional-CNN (1D-CNN) and
bidimensional-CNN (2D-CNN) to capture frequency and time
information was implemented in~\cite{vasquez2021transfer,quan2022end}. Other works have 
focused on representations that aim to model acoustic cues of PD;
in~\cite{vasquez2018towards,liu2022automatic}, authors used different feature 
sets to model different speech dimensions such as phonation, prosody, and 
articulation, and then these feature sets are used to classify PD patients and 
HCs. A similar approach was proposed in~\cite{klumpp2022phonetic}, where 
the authors modeled the phoneme articulation precision in PD patients using 
phonetic information. In~\cite{garcia2021cognitive}, prosody, articulation, and 
phonemic information features were combined to classify PD patients and HCs, 
showing that phonemic information was useful in discriminating PD patients 
with cognitive impairment from control subjects in a retelling task. 
In~\cite{Narendra2021}, a traditional pipeline solution based on Support Vector Machines (SVMs) and 
a 1D-CNN end-to-end approach were contrasted and compared, showing that the 
latter attain better results. 
An exemplar-based sparse representation was investigated in~\cite{Reddy2023} 
to avoid the time-consuming training phase of traditional machine learning 
solutions and to have a representation more robust to redundancy and 
noise in the data. %
A multimodal solution based on linguistics and acoustics cues was proposed in~\cite{escobargrisales23_interspeech} with promising results. %

Despite the growing attention within the speech community, the majority of
suggested solutions are typically evaluated under optimal conditions, i.e., training 
and testing data are matched, and often recordings are collected in a 
sound-proof booth using a professional audio setting. 
A few recent works consider moderately challenging scenarios, e.g.,  \cite{KARAN2021101216, VEETIL2024107494}, where training and testing conditions 
are not aligned. Nonetheless, only sustained vowel phonations and isolated words 
were evaluated. Consequently, the experimental setup may not accurately 
reflect a real production scenario in which automatic speech analysis 
would occur, i.e., speech production is typically evaluated by 
neurologists in continuous speech \cite{Steckhan2022}. This work seeks to overcome these limitations 
in experimentation by advocating for the exploration of foundational models 
and speech enhancement (SE) techniques to improve PD detection from speech 
in a real-world evaluation scenario. First, we demonstrate that foundational-based 
Parkinson's detectors are a viable solution in controlled conditions like those
in the standard PC-GITA~\cite{PC-GITA2014} corpus ({\bf s-PG-GITA}), which recordings were
collected in ideal conditions. 
Next, we show that foundational-based models trained on s-PC-GITA generalize better than both pipeline 
techniques~\cite{garcia2021cognitive} and end-to-end convolutional 
solutions~\cite{Narendra2021} when tested on realistic conditions provided with 
the extended PC-GITA~\cite{KARAN2021101216} ({\bf e-PC-GITA}) dataset, which recordings were 
collected in a real-world operating environment. Finally, we investigated the effect of 
off-the-shelf SE techniques\footnote{In this work, we use the term speech enhancement in a broad sense to indicate any pre-processing step that  improves the quality of the speech recordings and not limited to reduction of the background additive noise.} on the automatic analysis of Parkinson's speech,  where the goal is to align non-ideal testing recordings to the ideal training 
ones. Specifically, we process the e-PC-GITA recordings by performing 
the following three steps: (i) voice activity detection (VAD), (ii) speech 
dereverberation, and (iii) speech denoising. 
Experimental evidence demonstrates that speech enhancement is not only 
beneficial to deploy robust foundational-based PD detectors but 
also helps balancing sensitivity and specificity of the best self-supervised models.
Finally, we significantly boosted the PD  results on  the enhanced e-PC-GITA corpus by combining the two best foundational-based 
models, i.e., WavLM Base and Hubert Base, trained on s-PC-GITA data. To the best of the authors' knowledge, this is the first work assessing PD detection from speech in a real-world operating environment.%

\section{Proposed Foundational-based PD Detector  \& Speech Enhancement}
\label{sec2}

\begin{figure}[t]
    \centering
    \centerline{\includegraphics[width=8.6cm, height=5.1cm]{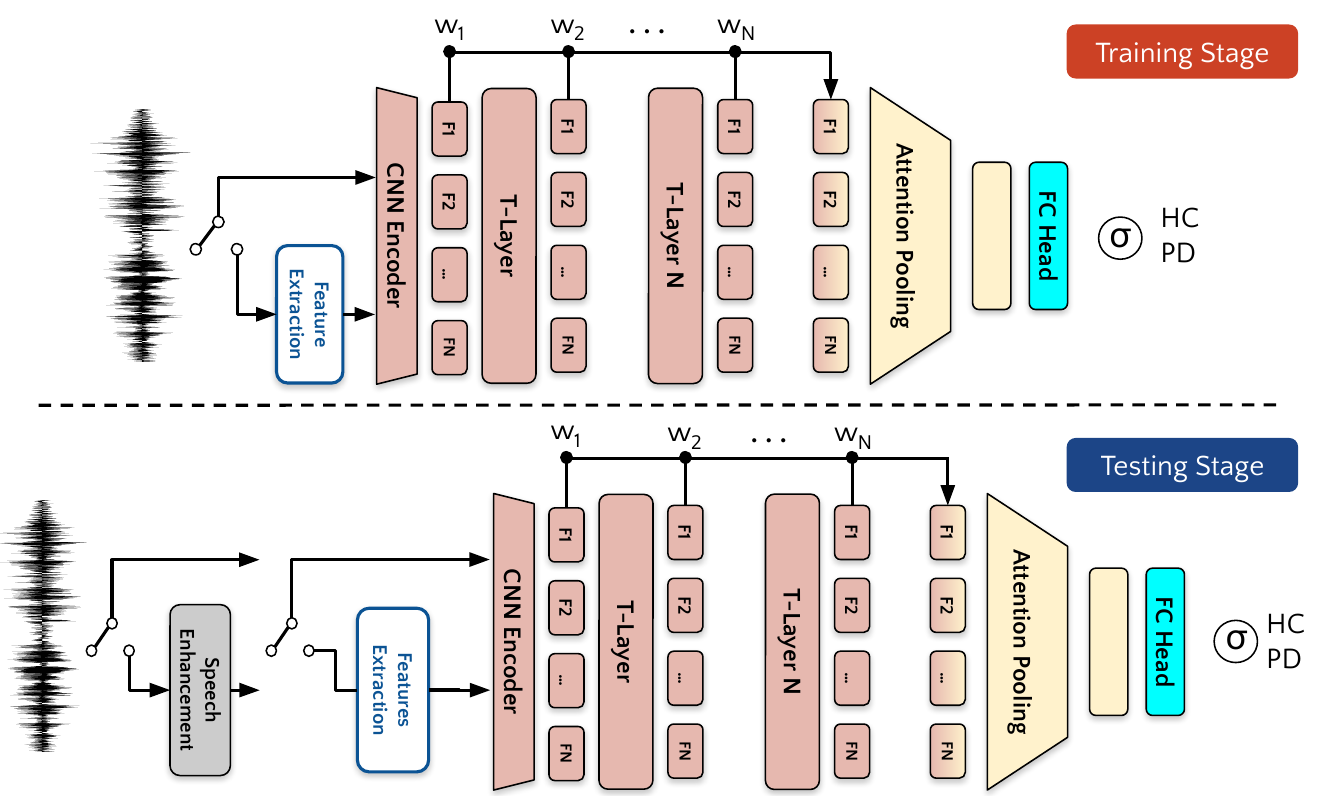}}
    \caption{Proposed foundational-based solution. Training in the upper panel, and testing in the bottom panel. Feature extraction is activated for SL models. At testing time e-PC-GITA data can be optionally enhanced; whereas training uses clean s-PC-GITA data only. The single-node output layer uses a Sigmoid. $F$ blocks indicate speech frames.}
    \label{fig:architecture}
    \vspace{-3mm}
\end{figure}
In Figure \ref{fig:architecture}, the proposed foundational-based model is depicted in the training (top panel), and testing (bottom panel) stages.  In training,  only s-PC-GITA data and labels are used. When testing on e-PC-GITA, recordings can be optionally enhanced (see Section \ref{sec:SE}). Models evaluated in this work are based on the Transformer architecture, and both self-supervised learning (SSL) and (weakly) supervised learning (SL) pre-trained models are investigated for the task of PD detection from speech. SSL models are effective in learning representations from  large amounts of unlabelled data~\cite{chen2022wavlm,babu2021xls, baevski2020wav2vec}, but SL-based foundational models have also recently shown to learn generalizable representations from large amounts of curated data~\cite{radford2023robust, gong23d_interspeech}.

\subsection{PD Detector}
\label{sec:PD_Model}
Foundational pre-trained encoders represent the backbone of the proposed 
PD detection model and serve to extract high-level representations from the 
input speech data. Through a weighted sum operation over frame representations 
across various layers, the backbone dynamically adjusts the importance of 
features extracted at different levels of abstraction, enhancing the model's 
ability to capture relevant information.
The foundational encoder is followed up by an attention pooling head to aggregate 
the frame-level representations into a single vector representation. 
This component employs learnable attention scores to focus on specific frames of 
the input, enabling the model to prioritize relevant information for the 
detection task.
After attention pooling, two fully connected linear layers with ReLU activation
functions are used to facilitate feature adaptation and transformation, enabling 
the model to refine its representations and capture discriminative features from 
the input speech data. The single-node output layer uses a Sigmoid activation function.
Our PD model is trained end-to-end, so that 
both the fully connected layers and the backbone itself are adapted 
to the final task. Hence the model learns task-specific features and optimizes 
its performance for PD detection\footnote{Code and pre-trained models are available at: \\ \url{https://github.com/K-STMLab/SSL4PR/}}.

\subsubsection{Foundational Backbone}
\label{sec:Foundational}

\noindent \textbf{SSL Models} The employed SSL models  operate directly on raw 
waveform and consist of an encoder network, which includes a CNN followed by a 
stack of transformer layers. Three different SSL solutions were considered: 
Wav2Vec2~\cite{baevski2020wav2vec}, HuBERT~\cite{hsu2021hubert}, and 
WavLM~\cite{chen2022wavlm}. Despite sharing a similar underlying architecture, 
these models differ in their pre-training objectives, offering distinct approaches 
to representation learning.
Wav2Vec2 utilizes a contrastive pre-training task on frame representations 
obtained by masking input audio, focusing on learning discriminative features.
HuBERT employs a similar strategy but also leverages MFCC features and clustering 
to create target representations for the contrastive task.
This approach is designed to learn more robust representations by leveraging the complementary information provided by MFCCs.
WavLM introduces masked speech denoising and prediction tasks during pre-training,
enhancing the model's robustness to noise and variations in the input data.
By considering these diverse pre-training tasks, we aim to explore the effectiveness 
of different strategies for learning representations from speech data and their 
impact on PD detection.
Finally, we also investigate the effects of multilingual pre-training and test the larger version of WavLM, and XLSR~\cite{babu2021xls}, both pre-trained on 
multilingual data. XLSR is a larger variant of Wav2Vec2.

\noindent \textbf{SL models}: We consider Whisper~\cite{radford2023robust}, a 
family of models that have been pre-trained on large-scale, curated datasets.
Whisper is an encoder-decoder model designed to process log-mel spectrograms 
of speech signals and is trained end-to-end in a supervised manner to 
generate text transcriptions.
Since our goal is PD detection, we focus solely on the encoder component of 
the Whisper model. The Whisper encoder comprises a CNN block that aggregates 
frame-level representations, specifically derived from log-mel representations, 
and a stack of transformer layers. 
To ensure a fair comparison in terms of model capacity, we consider the small 
version of Whisper, namely Whisper Small. %
The encoder component of this model has a comparable number of parameters with Wav2Vec2, and HuBERT, easing a fair comparison between 
SSL and SL models.

\subsection{Speech Enhancement Module}
\label{sec:SE}
Audio recordings collected in real-world conditions, e.g.,  patients visiting the clinic for a screening, usually contain background noise, reverberation,  and interfering speech - the interviewer suggests with quiet voice the next word to be spoken. To handle different real-world acoustic conditions and improve the robustness of our PD detectors, we can leverage speech enhancement techniques. 
In particular, we first removed the background 
interviewer interfering speech using rVADfast, an unsupervised segment-based
voice activity detector (VAD) presented in~\cite{TAN20201}. 
Next, we used a diffusion-based model for dereverberation~ \cite{welker22_interspeech}.
Finally, MP-SENet~\cite{lu23e_interspeech} is used to mitigate background noise. Speech enhancement takes place only in the testing phase, as shown in the bottom panel in Figure \ref{fig:architecture}.

\begin{table*}[h]
\centering
\caption{PD results averaged over 10-folds on s-PC-GITA. Mean value and standard deviation are reported.}
\label{tab:matched-conditions-res}
\begin{tabular}{@{}lllllll@{}}
\toprule
Model                                      & Accuracy                & F1-score                      & ROC-AUC                 & Sensitivity              & Specificity              \\ \midrule
1D-CNN \cite{Narendra2021}                 & 64.48 $\pm$ 12.03 & 60.15 $\pm$ 15.88 & 64.49 $\pm$ 12.02 & 53.67 $\pm$ 31.63 & 75.32 $\pm$ 30.15 \\
Phonemic \cite{escobargrisales23_interspeech}  & 68.76 $\pm$ 8.80  & 68.52 $\pm$ 8.85  & 68.77 $\pm$ 8.80  & 65.33 $\pm$ 10.69 & 72.20 $\pm$ 13.74 \\
Whisper Small \cite{radford2023robust}     & 78.04 $\pm$ 11.22 & 77.89 $\pm$ 11.34 & 78.04 $\pm$ 11.21 & 77.11 $\pm$ 13.32 & 78.98 $\pm$ 13.60 \\
Wav2Vec2 Base \cite{baevski2020wav2vec}    & 80.04 $\pm$ 8.75  & 79.80 $\pm$ 8.90  & 80.05 $\pm$ 8.74  & 74.56 $\pm$ 12.20 & 85.54 $\pm$ 12.59 \\
XLSR Large \cite{babu2021xls}                    & 78.04 $\pm$ 9.25  & 77.83 $\pm$ 9.39  & 78.05 $\pm$ 9.24  & 73.33 $\pm$ 13.07 & 82.77 $\pm$ 11.42 \\
HuBERT Base \cite{hsu2021hubert}           & 81.15 $\pm$ 8.84  & 80.84 $\pm$ 9.17  & 81.16 $\pm$ 8.84  & 76.44 $\pm$ 12.46 & 85.88 $\pm$ 14.94 \\
WavLM Base \cite{chen2022wavlm}            & 82.21 $\pm$ 8.17  & 81.99 $\pm$ 8.34  & 82.22 $\pm$ 8.17  & 75.22 $\pm$ 12.33 & 89.21 $\pm$ 9.22  \\ 
WavLM Large \cite{chen2022wavlm}           & 79.59 $\pm$ 9.14  & 79.15 $\pm$ 9.46  & 79.61 $\pm$ 9.13  & 70.78 $\pm$ 14.27 & 88.44 $\pm$ 13.12 \\ 
\bottomrule
\end{tabular}
\vspace{-2mm}
\end{table*}

\section{Experiments}
\label{sec3}
\subsection{Dataset}

{\bf Standard PC-GITA (s-PC-GITA)}: This dataset comprises samples collected from a total of 100 individuals, evenly divided 
into two groups: 50 with PD and 50 HC. There are 25 men and 25 women in 
each group. The patients were diagnosed by a neurologist. The healthy 
controls are free of any reported PD symptoms or other neurodegenerative 
disorder. The speaker's age is also matched and varies from 31 to 86 years old. 
The recordings were performed in a sound-proof booth at Clínica Noel of 
Medell\'{i}n in Colombia. %
The recording's sampling frequency is 44.1 KHz with a 16-bit resolution. All speech signals were downsampled to 16 kHz in this work, as in \cite{Narendra2021}. %
More information about the dataset can be found in  \cite{PC-GITA2014}.

\noindent{\bf Extended PC-GITA (e-PC-GITA)}:  It consists of recordings from 20 PD patients and 20 HC. As in 
the case of s-PC-GITA, age and gender are matched. 
The speakers were recorded in the Parkinson’s foundation of Medell\'{i}n. 
In this case, however, the acoustic conditions were not ideal, since recordings were collected 
in a real-world operating scenarios. Recordings were sampled at 16 KHz with 
a 16 bit-resolution. Patients were evaluated by the same neurologist who  performed the evaluations in the s-PC-GITA. 
e-PG-GITA is used as an independent dataset to evaluate PD solutions under realistic conditions, i.e., speakers simulate patients visiting the clinic for a screening.

The speech tasks from both datasets considered in this work are:  DDK exercises\footnote{Repetition of the sequence of syllables: /pa-ta-ka/, /pe-ta-ka/, /pa-ka-ta/, /pa/, /ka/, /ta/.}, read sentences, and monologues.

\begin{table*}[]
\centering

\caption{In the upper part, results original e-PC-GITA  are shown. In the lower part, results on enhanced e-PC-GITA  are given. The average performance of the models trained on s-PC-GITA across individual folds and their standard deviations are reported.}
\label{tab:unmatched-conditions-res}
\begin{tabular}{@{}llllll@{}}
\toprule
Model & Accuracy & F1-score & ROC-AUC & Sensitivity & Specificity \\ \midrule
\multicolumn{6}{c}{Original e-PC-GITA Recordings} \\ \midrule
1D-CNN \cite{Narendra2021} & 64.17 $\pm$ 10.85 & 59.64 $\pm$ 16.95 & 64.17 $\pm$ 10.85 & 84.83 $\pm$ 9.53 & 43.50 $\pm$ 27.76 \\
Phonemic \cite{escobargrisales23_interspeech}        & 50.17 $\pm$ 1.38  & 46.47 $\pm$ 0.93  & 50.17 $\pm$ 1.38  & 76.33 $\pm$ 3.93 & 24.00 $\pm$ 1.70 \\
Whisper Small \cite{radford2023robust} & 74.83 $\pm$ 5.65 & 73.83 $\pm$ 7.21 & 74.83 $\pm$ 5.65 & 87.33 $\pm$ 6.67 & 62.33 $\pm$ 16.55 \\
Wav2Vec2 Base \cite{baevski2020wav2vec} & 67.75 $\pm$ 4.33 & 67.17 $\pm$ 4.60 & 67.75 $\pm$ 4.33 & 77.00 $\pm$ 6.57 & 58.50 $\pm$ 12.42 \\
HuBERT Base \cite{hsu2021hubert} & 67.33 $\pm$ 7.95 & 64.10 $\pm$ 10.37 & 67.33 $\pm$ 7.95 & 92.00 $\pm$ 4.64 & 42.67 $\pm$ 19.89 \\
WavLM Base \cite{chen2022wavlm} & 68.58 $\pm$ 5.96 & 65.98 $\pm$ 7.65 & 68.58 $\pm$ 5.96 & 90.50 $\pm$ 9.25 & 46.67 $\pm$ 19.54 \\ \midrule
\multicolumn{6}{c}{Enhanced  e-PC-GITA Recordings} \\ \midrule
1D-CNN \cite{Narendra2021} & 57.00 $\pm$ 7.70 & 50.50 $\pm$ 13.28 & 57.00 $\pm$ 7.70 & 60.67 $\pm$ 31.81 & 53.33 $\pm$ 35.06 \\
Phonemic \cite{escobargrisales23_interspeech}        & 69.50 $\pm$ 2.36  & 69.40 $\pm$ 2.37 & 69.50 $\pm$ 2.36 & 66.00 $\pm$ 4.90 & 73.00 $\pm$ 5.52 \\
Whisper Small \cite{radford2023robust} & 76.25 $\pm$ 4.51 & 75.70 $\pm$ 5.52 & 76.25 $\pm$ 4.51 & 80.50 $\pm$ 8.63 & 72.00 $\pm$ 16.22 \\
Wav2Vec2 Base \cite{baevski2020wav2vec} & 76.75 $\pm$ 2.80 & 75.80 $\pm$ 3.09 & 76.75 $\pm$ 2.80 & 57.33 $\pm$ 5.07 & 96.17 $\pm$ 1.50 \\
HuBERT Base \cite{hsu2021hubert} & 82.75 $\pm$ 1.94 & 82.68 $\pm$ 1.96 & 82.75 $\pm$ 1.94 & 81.50 $\pm$ 5.45 & 84.00 $\pm$ 7.35 \\
WavLM Base \cite{chen2022wavlm} & 83.33 $\pm$ 3.48 & 83.13 $\pm$ 3.66 & 83.33 $\pm$ 3.48 & 77.17 $\pm$ 9.04 & 89.50 $\pm$ 8.53 \\ \bottomrule
\end{tabular}
\vspace{-3mm}
\end{table*}

\begin{table}[]
\centering
\caption{Results with Hubert Base (H) and WavLM Base (H) trained on all s-PC-GITA data and tested on enhanced e-PC-GITA.  H+W indicates system combination.}
\vspace{-1mm}
\label{tab:unmatched-conditions-res-all}
\resizebox{\columnwidth}{!}{%
\begin{tabular}{@{}llllll@{}}
\toprule
Model & Accuracy & F1-score & ROC-AUC & Sensitivity & Specificity \\ \midrule
HuBERT Base (H) & 86.67 & 86.63 & 86.67 & 81.67 & 91.67 \\
WavLM  Base (W)  & 85.00 & 85.00 & 85.00 & 83.33 & 86.67 \\
H + W & 88.33 & 88.32 & 88.33 & 85.00 & 91.67 \\ \bottomrule
\end{tabular}%
}
\vspace{-3.5mm}
\end{table}

\subsection{Experimental Setup}

All PD detection models studied in this work are trained on s-PC-GITA 
using a 10-fold speaker-independent cross-validation approach  
following~\cite{Narendra2021} if not stated otherwise. 
In each fold, speakers are partitioned such that a balanced set of speakers 
both in terms of gender and class is allocated for testing, while the 
remaining speakers are utilized for training. Every speaker 
was used only once for testing. Furthermore, the same speaker was not used in both, training 
and testing.
The foundational-based models were trained and evaluated using audio samples having a
fixed duration of 10 seconds, with silence padding applied when needed.

Foundational-based models are initialized with pre-trained weights and fine-tuned end-to-end for the 
PD detection task. 
Base and the Whisper-based models have $\sim$95M parameters; whereas, large models have $\sim$300M parameters.
We employed the Adam optimizer~\cite{adam_opt} with a learning rate of $10^{-4}$, 
which is warmup-decayed linearly throughout training, with the warmup phase 
comprising 10\% of the total training steps. 
The batch size is set to 32, and training continues for up to 10 epochs.
Performance is evaluated using accuracy, F1-score, ROC-AUC, sensitivity, and specificity. Results are presented as mean values and standard deviations 
across the 10 folds of cross-validation.
Fine-tuning and evaluation procedures are conducted on a single Nvidia\textsuperscript{\textregistered} A100 GPU with 80GB of memory. This setup facilitates the efficient exploration of various pre-trained models within a 
standardized experimental framework.
To assess the effectiveness of the proposed solution in real operative conditions, 
all models are also tested on the e-PC-GITA dataset. 
Finally, we investigated the effect of the SE techniques, 
discussed in Section~\ref{sec:SE}, on the e-PC-GITA data.

\noindent\textbf{Baselines}: To better assess the proposed foundational-based solutions, we have built two baseline systems, which were proposed to tackle PD detection using the s-PC-GITA data. The first model leverages a 1-dimensional CNN (1D-CNN) as 
proposed in~\cite{Narendra2021}. It serves as a comparison against 
other deep models.
The 1D-CNN model has 60K trainable parameters and was deployed following the protocol outlined in the original paper~\cite{Narendra2021} for training, validation, and testing\footnote{Note that \cite{Narendra2021} did not provide the definition of the folds, and that might explain differences in the standard deviations with our results.}.
The second model employs an SVM-based classifier trained on phonemic 
features \cite{escobargrisales23_interspeech}, which are based on the 
posterior probability of a speech frame belonging to one (or more) of 
18 phonological classes~\cite{vasquezcorrea19_interspeech}. This model has 
demonstrated high efficacy in addressing the PD detection 
task~\cite{escobargrisales23_interspeech}. 

\subsection{Results}

\noindent\textbf{Ideal Testing Conditions - s-PC-GITA}:  Table~\ref{tab:matched-conditions-res} presents experimental results obtained under matched training and testing conditions. 
Foundational-based models consistently outperform baseline solutions, with the best-performing foundational model, WavLM Base, achieving accuracy, F1-score, and ROC-AUC of  82.21\%, 81.99\%, and 82.22\%, respectively.
Interestingly, SSL solutions demonstrate better performance compared to the SL 
model in most cases, underscoring the superior capabilities 
of SSL-based features. The SL model trained to assess specific tasks  may  tend to eliminate information available in the input speech potentially limiting their effectiveness
Another interesting outcome in Table~\ref{tab:matched-conditions-res} is that 
base models outperform larger multilingual models (roughly +2\% in terms 
of average accuracy), indicating that multilingual pre-training may not 
significantly benefit the PD detection task.
Given the limited amount of training data, larger models may also be 
more prone to over-fitting, potentially resulting in decreased overall 
performance.
In sum, foundational-based models outperform both CNN-based and SVM-based 
baselines under ideal testing conditions, indicating the effectiveness 
of leveraging pre-trained models for PD detection. While SSL models 
generally exhibit a better performance, the impact of multilingual 
pre-training on performance remains limited, suggesting careful consideration 
in model selection. %

\noindent \textbf{Real-World Testing Conditions - e-PC-GITA:}
In evaluating the generalization capability of the foundational-based models presented above, we assess their performance on the e-PC-GITA dataset.
Table~\ref{tab:unmatched-conditions-res} reports the average performance 
of the models trained through the 10-fold cross-validation scheme on the 
s-PC-GITA dataset and tested (without re-training) on the e-PC-GITA data.
The upper part of Table~\ref{tab:unmatched-conditions-res} shows results on the 
original e-PC-GITA data; it can be observed a substantial reduction in performance 
across all models when compared to the ideal testing conditions on 
s-PC-GITA. However, foundational-based models still consistently outperform 
both CNN- and SVM-based baselines.\footnote{In s-PC-GITA tests, variability comes from both model parameters, and testing data. In e-PC-GITA, variability is only from the model parameters. That may explain  the lower standard deviations despite more challenging testing conditions represented by e-PC-GITA.}
Notably,  SSL-based models show a more pronounced performance drop than
Whisper Small. 
This may be due to the inherent multi-conditional training of the SL model, 
trained on  diverse datasets covering a broad distribution of audio. 

The latter observation convinced us that we could boost the performance of 
the studied models, especially SSL-based solutions, by improving the 
quality of the testing recording to better resemble the ideal training 
conditions. A visual inspection of the lower part of 
Table~\ref{tab:unmatched-conditions-res}, which lists the experimental 
results on the enhanced e-PC-GITA data, confirms our intuition. 
Foundational-based solutions, particularly SSL-based ones, outperform the two
baselines systems. Moreover,  HuBERT and WavLM Base, i.e., two SSL-based models,
achieve the highest performance, with accuracy, F1-score, and ROC of 
83.33\%, 83.13\%, and 83.33\%, respectively, and they also show  more balanced sensitivity and specificity.
Interestingly, while the foundational-base models show considerable 
enhancement, the 1D-CNN model exhibits a notable decrease in performance. 

\noindent \textbf{System Combination}:
In the above experiments on e-PC-GITA, we have used the 
10 models built on the 10-fold s-PC-GITA in order to ease the 
comparison between ideal and real-world operating conditions. Nonetheless, better
models can be deployed by using all s-PC-GITA data to train a single 
model. We thus retrained the best-performing models, namely HuBERT Base 
(H) and WavLM Base (W), using s-PC-GITA dataset for training 
and validation, and assessed these two models on the enhanced version of e-PC-GITA. By comparing the first two rows in Table~\ref{tab:unmatched-conditions-res-all} with the corresponding entries in Table~\ref{tab:unmatched-conditions-res}, we can confirm our expectations. 
Finally, combining the predictions of these two models at an inference time using a simple averaging approach resulted in a system combination (H + W) that 
outperformed individual models, achieving an accuracy, F1-score, and ROC-AUC of around 88.33\%, 88.13\%, and 88.23\%, respectively. %
\vspace{-1.8mm}
\section{Conclusion}
\label{sec4}
In this work, we have first shown that foundational models can be effectively 
fine-tuned to the PD detection task, outperforming conventional
models previously reported in the literature. 
Next, we have shown that (weakly) supervised foundational models fine-tuned on 
s-PC-GITA attain the best results on unseen real-world conditions, represented 
by e-PC-GITA. We have speculated that this is caused by the 
inherent Whisper multi-condition training. Nonetheless, a more severe drop in performance was
observed for all other evaluated models. Therefore, we used off-the-shelf speech enhancement techniques to try to align training and testing conditions. As expected,
foundational-based models improved significantly, and Hubert Base and WavLM 
Base, which are SSL-based models, reported the best results on the enhanced e-PC-GITA. 
Finally,  a simple system combination of Huber Base and WavLM Base significantly boosted the 
PD detection results on the enhanced e-PC-GITA dataset.

\clearpage
\bibliographystyle{IEEEtran}
\balance
\bibliography{mybib}

\begin{thebibliography}{10}
\providecommand{\url}[1]{#1}
\csname url@samestyle\endcsname
\providecommand{\newblock}{\relax}
\providecommand{\bibinfo}[2]{#2}
\providecommand{\BIBentrySTDinterwordspacing}{\spaceskip=0pt\relax}
\providecommand{\BIBentryALTinterwordstretchfactor}{4}
\providecommand{\BIBentryALTinterwordspacing}{\spaceskip=\fontdimen2\font plus
\BIBentryALTinterwordstretchfactor\fontdimen3\font minus \fontdimen4\font\relax}
\providecommand{\BIBforeignlanguage}[2]{{%
\expandafter\ifx\csname l@#1\endcsname\relax
\typeout{** WARNING: IEEEtran.bst: No hyphenation pattern has been}%
\typeout{** loaded for the language `#1'. Using the pattern for}%
\typeout{** the default language instead.}%
\else
\language=\csname l@#1\endcsname
\fi
#2}}
\providecommand{\BIBdecl}{\relax}
\BIBdecl

\bibitem{hornykiewicz1998biochemical}
O.~Hornykiewicz, ``Biochemical aspects of {P}arkinson's disease,'' \emph{Neurology}, vol.~51, no. 2 Suppl 2, pp. S2--S9, 1998.

\bibitem{Poewe2017}
W.~Poewe \emph{et~al.}, ``Parkinson disease,'' \emph{Nature Rev. Dis. Primers}, vol.~23, 2017.

\bibitem{Rijk2000}
M.~C. de~Rijk \emph{et~al.}, ``Prevalence of parkinson’s disease in europe: A collaborative study of population-based cohorts. neurologic diseases in the elderly research group,'' \emph{Neurology}, vol.~54, pp. S21--S23, 2020.

\bibitem{mu2017parkinson}
J.~Mu \emph{et~al.}, ``Parkinson's disease subtypes identified from cluster analysis of motor and non-motor symptoms,'' \emph{Frontiers in aging neuroscience}, vol.~9, p. 301, 2017.

\bibitem{Jankovic2008}
J.~Jankovic, ``Parkinson’s disease: Clinical features and diagnosis,'' \emph{J. Neural. Neurosurg. Psychiatry}, vol.~79, pp. 368--376, 2008.

\bibitem{pinto2004treatments}
S.~Pinto \emph{et~al.}, ``Treatments for dysarthria in parkinson's disease,'' \emph{The Lancet Neurology}, vol.~3, no.~9, pp. 547--556, 2004.

\bibitem{Rusz2013}
J.~Rusz \emph{et~al.}, ``{Imprecise vowel articulation as a potential early marker of Parkinson's disease: effect of speaking task},'' \emph{Journal of the Acoustical Society of America}, vol. 134, no.~3, pp. 2171--2181, 2013.

\bibitem{Rodriguez-Oroz2009}
M.~C. Rodriguez-Oroz \emph{et~al.}, ``Initial clinical manifestations of parkinson’s disease: Features and pathophysiological mechanisms,'' \emph{Lancet Neurol}, vol.~8, pp. 1128--1139, 2009.

\bibitem{sonawane2021speech}
B.~Sonawane and P.~Sharma, ``Speech-based solution to {P}arkinson’s disease management,'' \emph{Multimedia Tools and Applications}, vol.~80, no.~19, pp. 29\,437--29\,451, 2021.

\bibitem{rios2022end}
C.~Rios-Urrego \emph{et~al.}, ``End-to-end {P}arkinson’s disease detection using a deep convolutional recurrent network,'' in \emph{International Conference on Text, Speech, and Dialogue}.\hskip 1em plus 0.5em minus 0.4em\relax Springer, 2022, pp. 326--338.

\bibitem{vasquez2021transfer}
J.~V{\'a}squez-Correa \emph{et~al.}, ``Transfer learning helps to improve the accuracy to classify patients with different speech disorders in different languages,'' \emph{Pattern Recognition Letters}, vol. 150, pp. 272--279, 2021.

\bibitem{quan2022end}
C.~Quan \emph{et~al.}, ``End-to-end deep learning approach for {P}arkinson’s disease detection from speech signals,'' \emph{Biocybernetics and Biomedical Engineering}, vol.~42, no.~2, pp. 556--574, 2022.

\bibitem{vasquez2018towards}
J.~C. V{\'a}squez-Correa \emph{et~al.}, ``Towards an automatic evaluation of the dysarthria level of patients with parkinson's disease,'' \emph{Journal of communication disorders}, vol.~76, pp. 21--36, 2018.

\bibitem{liu2022automatic}
Y.~Liu \emph{et~al.}, ``Automatic assessment of parkinson's disease using speech representations of phonation and articulation,'' \emph{IEEE/ACM Transactions on Audio, Speech, and Language Processing}, vol.~31, pp. 242--255, 2022.

\bibitem{klumpp2022phonetic}
P.~Klumpp \emph{et~al.}, ``The phonetic footprint of parkinson’s disease,'' \emph{Computer Speech \& Language}, vol.~72, p. 101321, 2022.

\bibitem{garcia2021cognitive}
A.~M. Garc{\'\i}a \emph{et~al.}, ``Cognitive determinants of dysarthria in parkinson's disease: an automated machine learning approach,'' \emph{Movement Disorders}, vol.~36, no.~12, pp. 2862--2873, 2021.

\bibitem{Narendra2021}
N.~Narendra, B.~Schuller, and P.~Alku, ``The detection of parkinson's disease from speech using voice source information,'' \emph{IEEE/ACM Transactions on Audio, Speech, and Language Processing}, vol.~29, pp. 1925--1936, 2021.

\bibitem{Reddy2023}
M.~K. Reddy and P.~Alku, ``Exemplar-based sparse representations for detection of parkinson's disease from speech,'' \emph{IEEE/ACM Transactions on Audio, Speech, and Language Processing}, vol.~31, pp. 1386--1396, 2023.

\bibitem{escobargrisales23_interspeech}
D.~Escobar-Grisales, T.~Arias-Vergara, C.~D. Ríos-Urrego, E.~Nöth, A.~M. García, and J.~R. Orozco-Arroyave, ``{An Automatic Multimodal Approach to Analyze Linguistic and Acoustic Cues on Parkinson's Disease Patients},'' in \emph{Proc. INTERSPEECH}, 2023, pp. 1703--1707.

\bibitem{KARAN2021101216}
B.~Karan, S.~S. Sahu, J.~R. Orozco-Arroyave, and K.~Mahto, ``Non-negative matrix factorization-based time-frequency feature extraction of voice signal for parkinson's disease prediction,'' \emph{Computer Speech \& Language}, vol.~69, 2021.

\bibitem{VEETIL2024107494}
``Robust language independent voice data driven parkinson’s disease detection,'' \emph{Engineering Applications of Artificial Intelligence}, vol. 129, 2024.

\bibitem{Steckhan2022}
P.~Hecker, N.~Steckhan, F.~Eyben, B.~Schuller, and B.~Arnrich, ``Voice analysis for neurological disorder recognition-a systematic review and perspective on emerging trends,'' \emph{Front Digit Health}, vol.~4, 2022.

\bibitem{PC-GITA2014}
J.~Orozco-Arroyave, J.~Arias-Londonõ, J.~Vargas-Bonilla, M.~Gonzalez-Rativa, and E.~Noth, ``{New Spanish speech corpus database for the analysis of people suffering from Parkinson’s disease},'' in \emph{Proc. LREC}, 2014, p. 342–347.

\bibitem{chen2022wavlm}
S.~Chen \emph{et~al.}, ``Wavlm: Large-scale self-supervised pre-training for full stack speech processing,'' \emph{IEEE Journal of Selected Topics in Signal Processing}, vol.~16, no.~6, pp. 1505--1518, 2022.

\bibitem{babu2021xls}
A.~Babu \emph{et~al.}, ``Xls-r: Self-supervised cross-lingual speech representation learning at scale,'' \emph{arXiv preprint arXiv:2111.09296}, 2021.

\bibitem{baevski2020wav2vec}
A.~Baevski \emph{et~al.}, ``Wav2vec 2.0: A framework for self-supervised learning of speech representations,'' \emph{Advances in Neural Information Processing Systems}, vol.~33, pp. 12\,449--12\,460, 2020.

\bibitem{radford2023robust}
A.~Radford \emph{et~al.}, ``Robust speech recognition via large-scale weak supervision,'' in \emph{International Conference on Machine Learning}.\hskip 1em plus 0.5em minus 0.4em\relax PMLR, 2023, pp. 28\,492--28\,518.

\bibitem{gong23d_interspeech}
Y.~Gong \emph{et~al.}, ``{Whisper-AT: Noise-Robust Automatic Speech Recognizers are Also Strong General Audio Event Taggers},'' in \emph{Proc. INTERSPEECH 2023}, 2023, pp. 2798--2802.

\bibitem{hsu2021hubert}
W.-N. Hsu \emph{et~al.}, ``Hubert: Self-supervised speech representation learning by masked prediction of hidden units,'' \emph{IEEE/ACM Transactions on Audio, Speech, and Language Processing}, vol.~29, pp. 3451--3460, 2021.

\bibitem{TAN20201}
Z.-H. Tan, A.~kr. Sarkar, and N.~Dehak, ``rvad: An unsupervised segment-based robust voice activity detection method,'' \emph{Computer Speech \& Language}, vol.~59, pp. 1--21, 2020.

\bibitem{welker22_interspeech}
S.~Welker, J.~Richter, and T.~Gerkmann, ``{Speech Enhancement with Score-Based Generative Models in the Complex STFT Domain},'' in \emph{Proc. Interspeech 2022}, 2022, pp. 2928--2932.

\bibitem{lu23e_interspeech}
Y.-X. Lu, Y.~Ai, and Z.-H. Ling, ``{MP-SENet: A Speech Enhancement Model with Parallel Denoising of Magnitude and Phase Spectra},'' in \emph{Proc. INTERSPEECH 2023}, 2023, pp. 3834--3838.

\bibitem{adam_opt}
I.~Loshchilov and F.~Hutter, ``Decoupled weight decay regularization,'' in \emph{International Conference on Learning Representations}, 2018.

\bibitem{vasquezcorrea19_interspeech}
J.~Vásquez-Correa \emph{et~al.}, ``{Phonet: A Tool Based on Gated Recurrent Neural Networks to Extract Phonological Posteriors from Speech},'' in \emph{Proc. Interspeech 2019}, 2019, pp. 549--553.

\end{thebibliography}

\end{document}